\documentclass[manuscript,screen]{acmart}

\AtBeginDocument{%
  }

\setcopyright{acmlicensed}
\copyrightyear{2018}
\acmYear{2018}
\acmDOI{XXXXXXX.XXXXXXX}

\acmConference[Conference acronym 'XX]{Make sure to enter the correct
  conference title from your rights confirmation emai}{June 03--05,
  2018}{Woodstock, NY}
\acmISBN{978-1-4503-XXXX-X/18/06}

\begin{document}

\title{From Text to Blueprint: Leveraging Text-to-Image Tools for Floor Plan Creation}

\author{Xiaoyu Li}
\author{Jonathan Benjamin}
\author{Xin Zhang}
\affiliation{%
  \institution{TAG, HUST}
  \country{.}
}


\renewcommand{\shortauthors}{Li et al.}


\begin{abstract}

Artificial intelligence is revolutionizing architecture through text-to-image synthesis, converting textual descriptions into detailed visual representations. We explore AI-assisted floor plan design, focusing on technical background, practical methods, and future directions. Using tools like, Stable Diffusion, AI leverages models such as Generative Adversarial Networks and Variational Autoencoders to generate complex and functional floorplans designs. We evaluate these AI models' effectiveness in generating residential floor plans from text prompts. Through experiments with reference images, text prompts, and sketches, we assess the strengths and limitations of current text-to-image technology in architectural visualization. Architects can use these AI tools to streamline design processes, create multiple design options, and enhance creativity and collaboration. We highlight AI's potential to drive smarter, more efficient floorplan design, contributing to ongoing discussions on AI integration in the design profession and its future impact.
\end{abstract}



\maketitle

\section{Motivation}

Artificial intelligence (AI) has become a transformative force in the field of architecture\cite{mcguire1983theory,groat2013architectural,garlan1994exploiting,machairas2014algorithms,demirbacs2003focus,caetano2020computational,xu2017taxonomy,kotnik2010digital,hollberg2016lca,aliakseyeu2006computer,akin1996frames,lomas2007architectural}, offering innovative solutions to design challenges and creative exploration, including 2D and 3D vision technology \cite{Rae2021ScalingLM,Xu2022GroupViTSS,li2023efficient,Thoppilan2022LaMDALM, Brown2020LanguageMA,Liu2022FewShotPF}. Among the many applications of AI in architecture, text-to-image synthesis\cite{brooks2023instructpix2pix,hertz2022prompt2prompt,li2023layerdiffusion,li2024tuning,meng2021sdedit,tumanyan2023plug} stands out as a promising approach for generating visual representations from textual descriptions. This technology uses advanced machine learning algorithms to convert written instructions into detailed and realistic images, opening up new possibilities for architectural visualization and design. This article will focus on the application of AI-assisted floor plan design, discussing its technical background, practical methods, and future development directions.

The emergence of text-to-image generation tools, such as Midjourney~\cite{Midjourney}, Stable Diffusion~\cite{stablediffusion}, and Dall-E~\cite{dalle2}, has revolutionized the way architects conceptualize and communicate their ideas. These tools leverage the power of deep learning models, including Generative Adversarial Networks (GANs)~\cite{GANs,shiftgan} and Variational Autoencoders (VAEs)~\cite{VQVAE}, to generate complex floor plans, building layouts, and spatial configurations based on text prompts. Trained on large datasets of text-image pairs, these models can capture the nuances of architectural design and produce visually striking outputs with remarkable accuracy.

Firstly, the role of GANs in text-to-image synthesis cannot be overlooked. Through adversarial training, GANs continuously improve the quality and accuracy of generated images. VAEs are also crucial tools in text-to-image synthesis. By encoding input data into a latent space and then decoding it to generate images, VAEs can capture complex features within the data. In architectural design, VAEs can encode different characteristics of building elements into the latent space, allowing for the consideration of these features' combinations and variations when generating floor plans. This ensures that the generated floor plans not only have high visual quality but also meet the functional and practical requirements of architectural design.

With the development of large models, significant potential has been shown in image generation and editing, video generation and editing~\cite{ho2022imagenvideo,videofusion}, and 3D content generation and editing~\cite{li2024art3d,magic3d,dreamfusion,diffusion3d}. Therefore, We aim to explore the capabilities of text-to-image models in generating residential floor plans. Through a comparative analysis of Stable Diffusion, we hope to evaluate the effectiveness of these AI systems in creating coherent and functional architectural designs from textual input~\cite{wang2024exploring}. By conducting experiments involving reference images, text prompts, and hand-drawn sketches, we seek to highlight the strengths and limitations of current text-to-image technology in the field of architectural visualization.

In practice, architects can use these AI tools to streamline the design process~\cite{li2024generating,li2023sketch,chaillou2020archigan,Para2021SketchGenGC,zhang2024boosting}. First, the architect inputs detailed text describing the floor plan, including the number, size, function, and relative positions of rooms. The AI tool then parses these text descriptions to generate initial floor plans. These floor plans can serve as a foundation for further modification and optimization by the architect, helping them complete design tasks more efficiently. Additionally, architects can use AI tools to generate multiple design options and select the one that best meets their needs, enhancing design flexibility and diversity.

Our research is motivated by the potential of AI-driven design tools to simplify the architectural design process, enhance creativity, and foster collaboration between architects, designers, and clients. By examining the outputs generated by these AI models and assessing their adherence to architectural principles and spatial logic, we hope to provide valuable insights into the evolving role of AI in architectural practice. Moreover, we aim to contribute to the ongoing discussion on the integration of AI technology in the design profession and its impact on future architectural innovation.

In conclusion, We aim to explore the intersection of artificial intelligence and architectural creativity, revealing the transformative potential of text-to-image models in floor plan generation. By investigating the capabilities of AI tools in architectural design and their impact on the design process, we hope to open new chapters in computational design and collaborative innovation in architecture. The advancement of AI technology not only provides architects with powerful design tools but also brings endless possibilities for the future of architectural design. Through continuous exploration and innovation, we believe AI will play an increasingly important role in architectural design, driving the industry towards a smarter and more efficient era.

\section{Methodology}

We explore the same objective as previous research: to verify if it is possible to use text-to-image technology, fine-tuned with a large collection of floor plan data, to obtain accurate floor layouts for a single-story, single-family home. Additionally, more robustness tests were conducted to draw broader and more useful conclusions about the potential of these tools. Our methodology followed these steps: First, we used a large language model~\cite{openai2023gpt4} to generate text prompts for residential floor plans. Then, we conducted experiments using the official stable diffusion v1.5 model~\cite{stablediffusion} weights and the weights we fine-tuned on a large number of real floor plans. This allowed us to obtain different floor plan generation results. These experiments verified that AI tools can provide designers with floor plans that can be further modified, enabling them to generate multiple design options and select the one that best meets their needs. This approach increases design flexibility and diversity.

Next, we will describe in detail the extensive text prompts generated through GPT.

\begin{figure*}[tbp]
    \centering
    \includegraphics[width=0.99\textwidth]{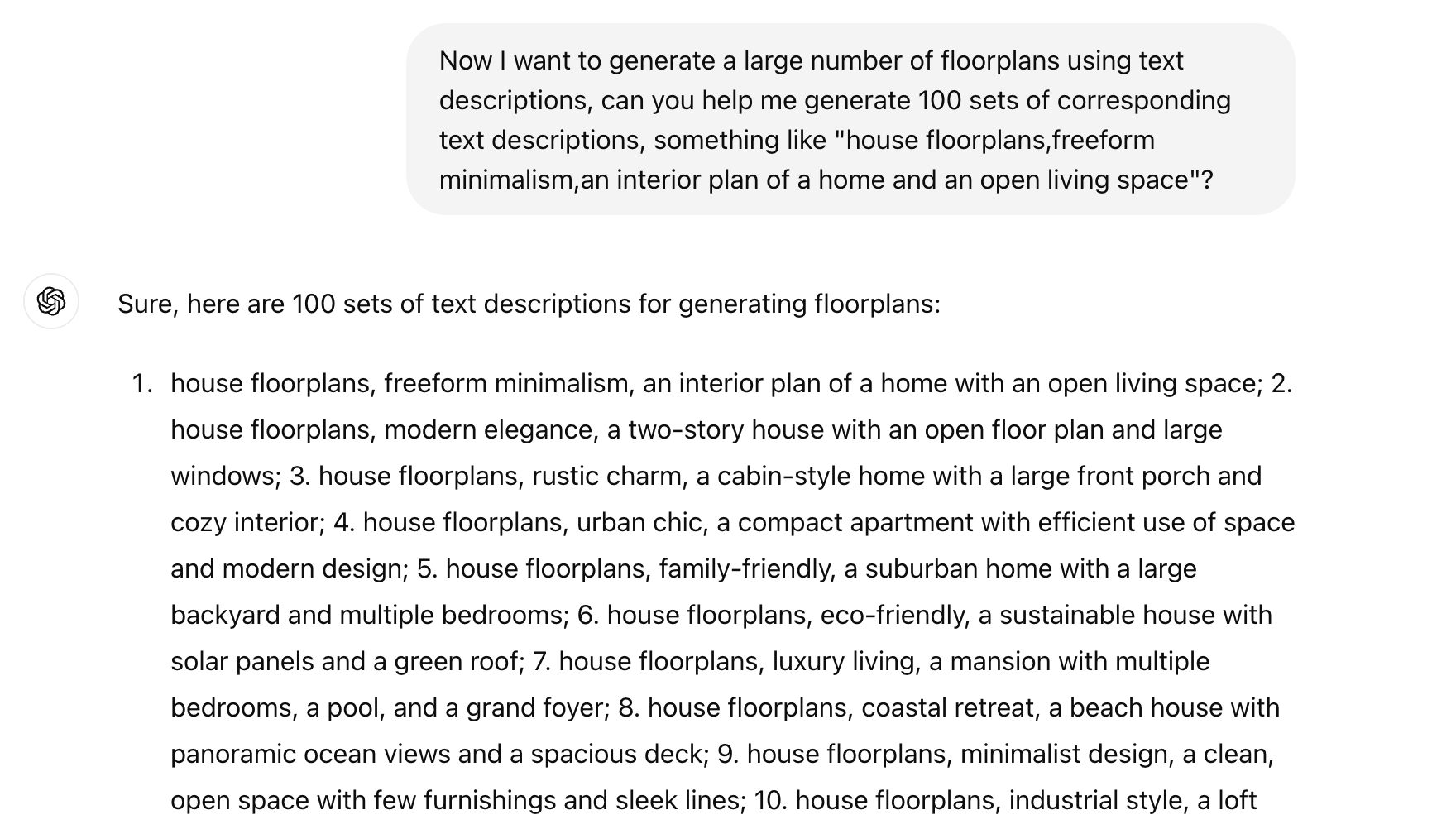}
    \caption{We demonstrate the process of using GPT to generate textual descriptions. }
\end{figure*}

\begin{figure*}[tbp]
    \centering
    \includegraphics[width=0.99\textwidth]{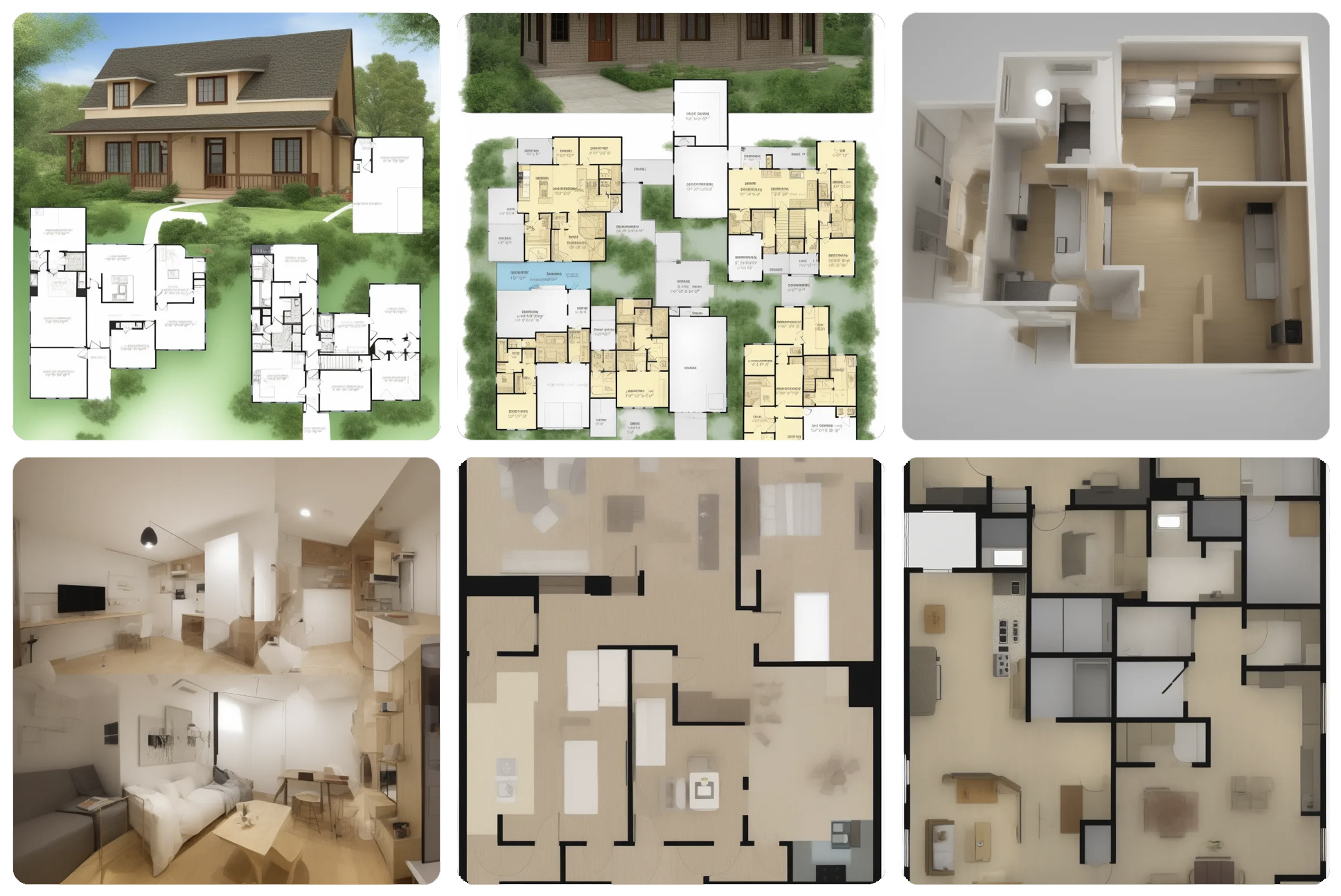}
    \caption{Floorplans-01: Generated by Stable Diffusion models. }
\end{figure*} 

\begin{figure*}[tbp]
    \centering
    \includegraphics[width=0.99\textwidth]{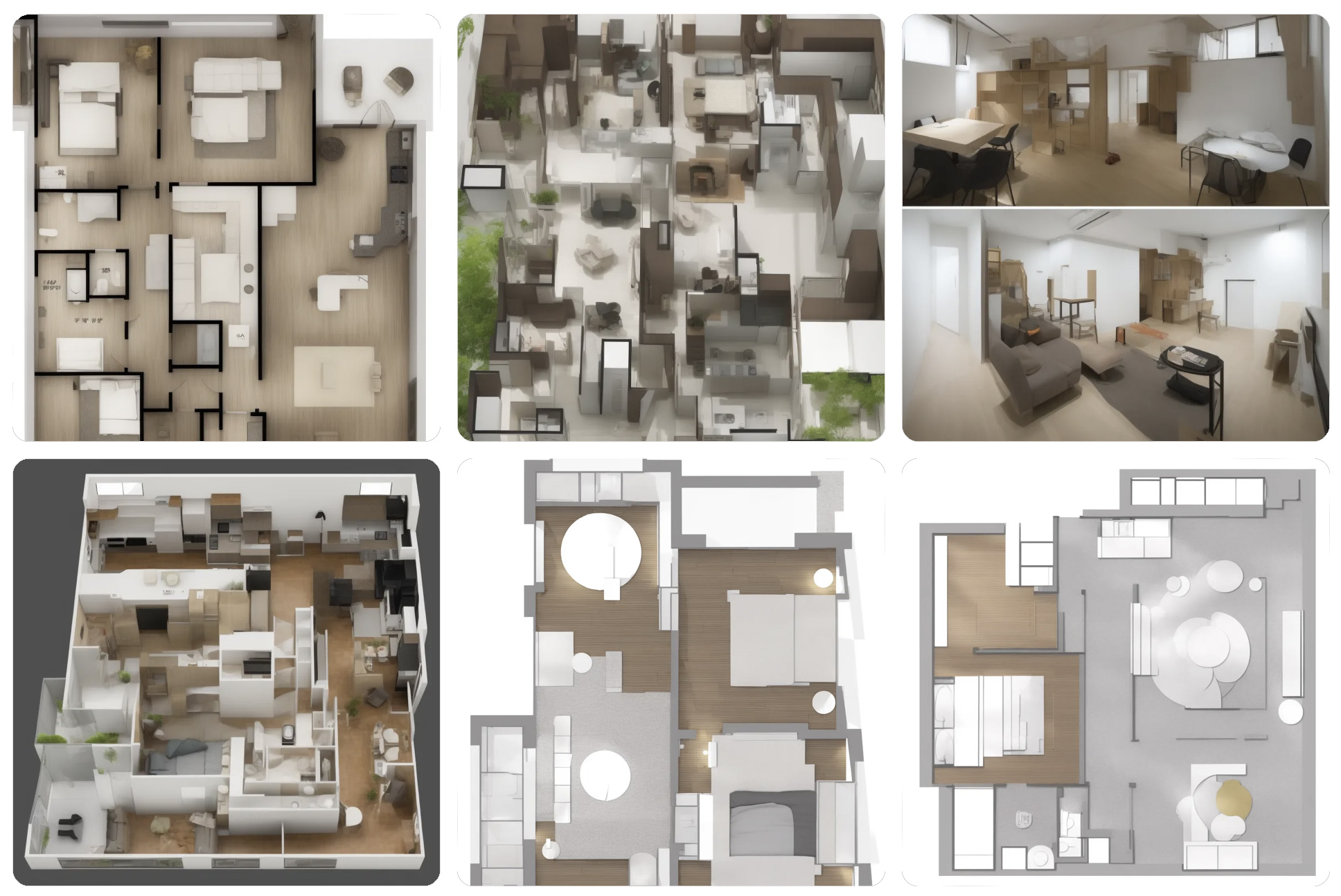}
    \caption{Floorplans-02: Generated by Stable Diffusion models. }
\end{figure*} 

\begin{figure*}[tbp]
    \centering
    \includegraphics[width=0.99\textwidth]{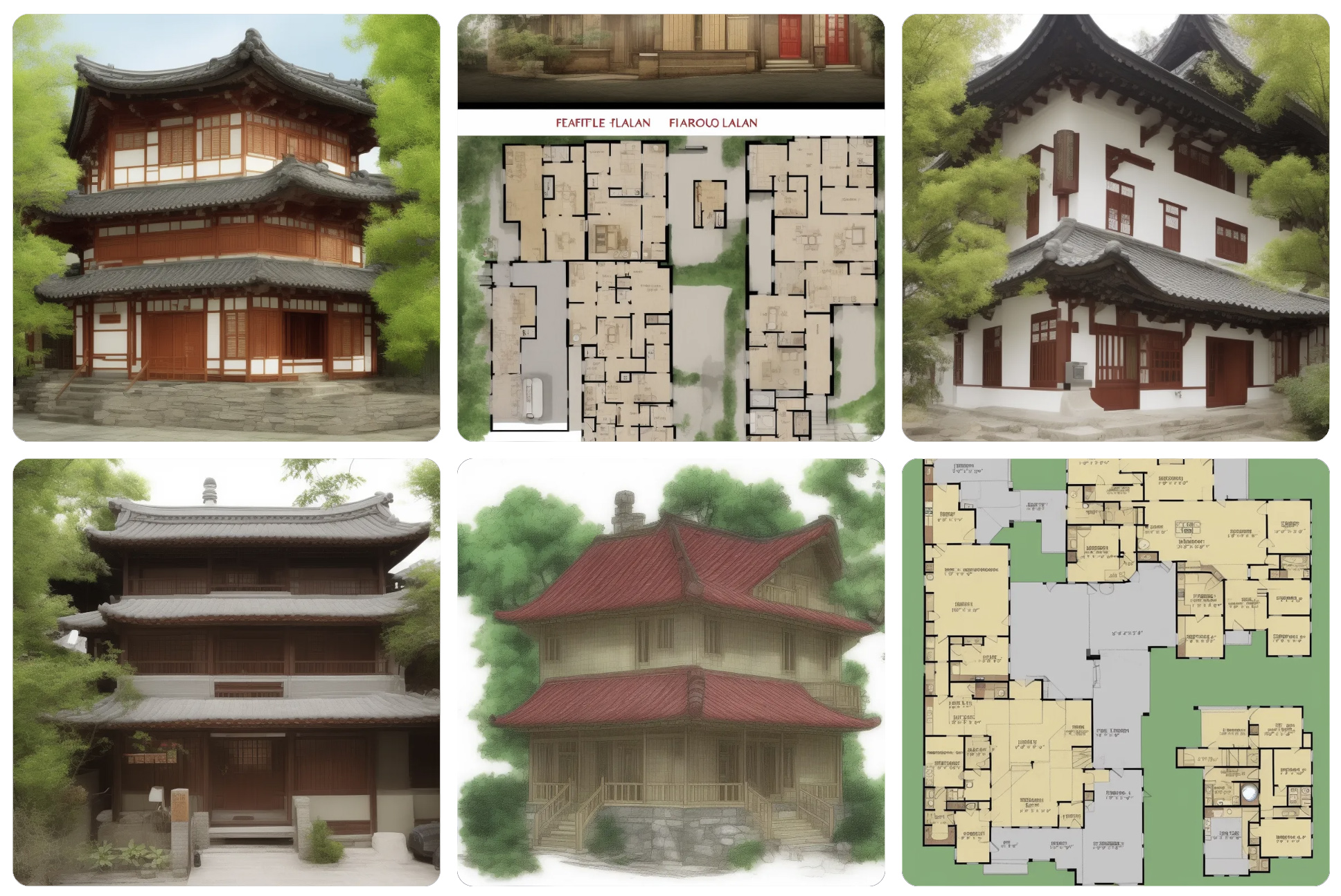}
    \caption{Floorplans-03: Generated by Stable Diffusion models. }
\end{figure*} 

\begin{figure*}[tbp]
    \centering
    \includegraphics[width=0.99\textwidth]{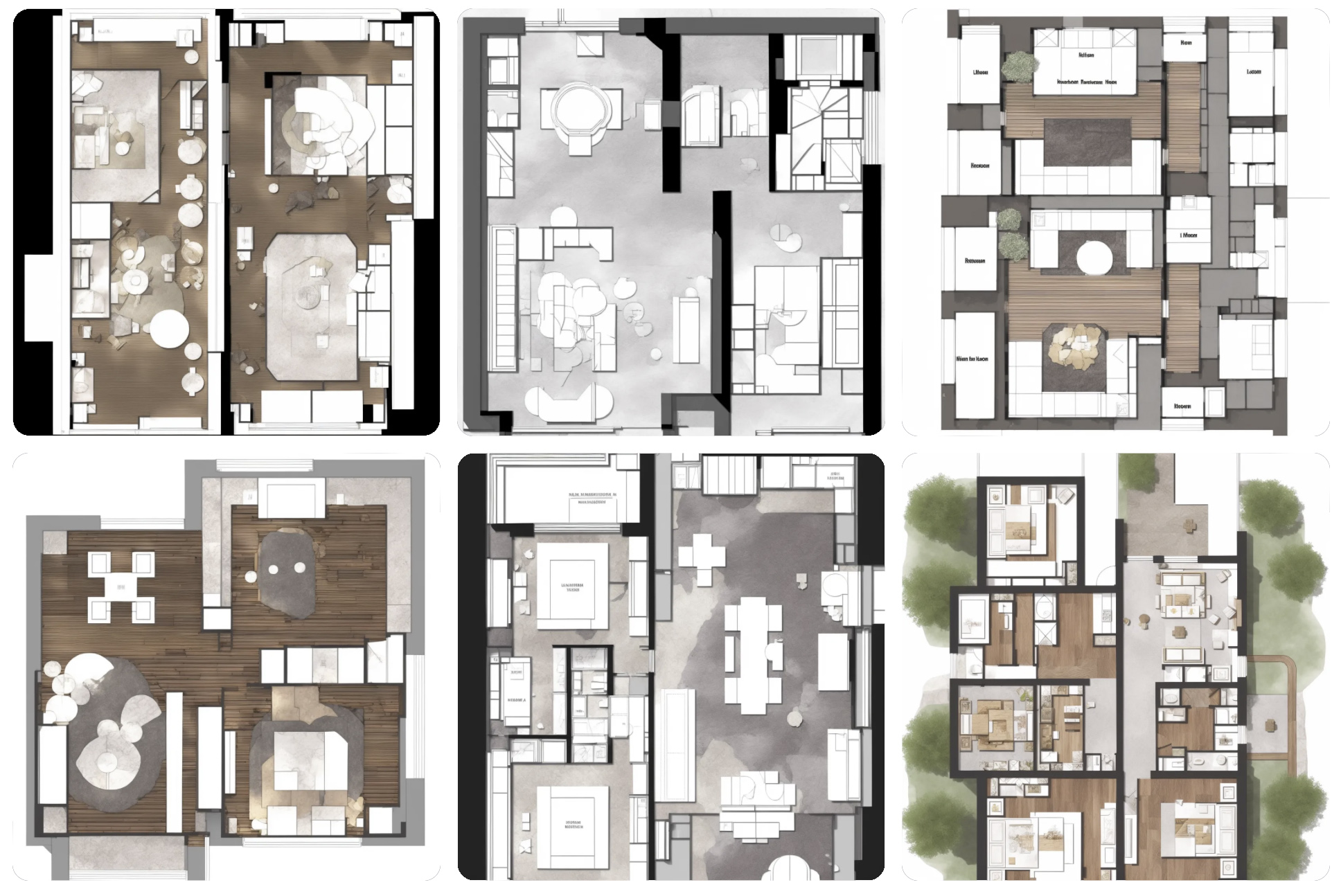}
    \caption{Floorplans-04: Generated by Fine-tuned LoRA weights. }
\end{figure*} 

\begin{figure*}[tbp]
    \centering
    \includegraphics[width=0.99\textwidth]{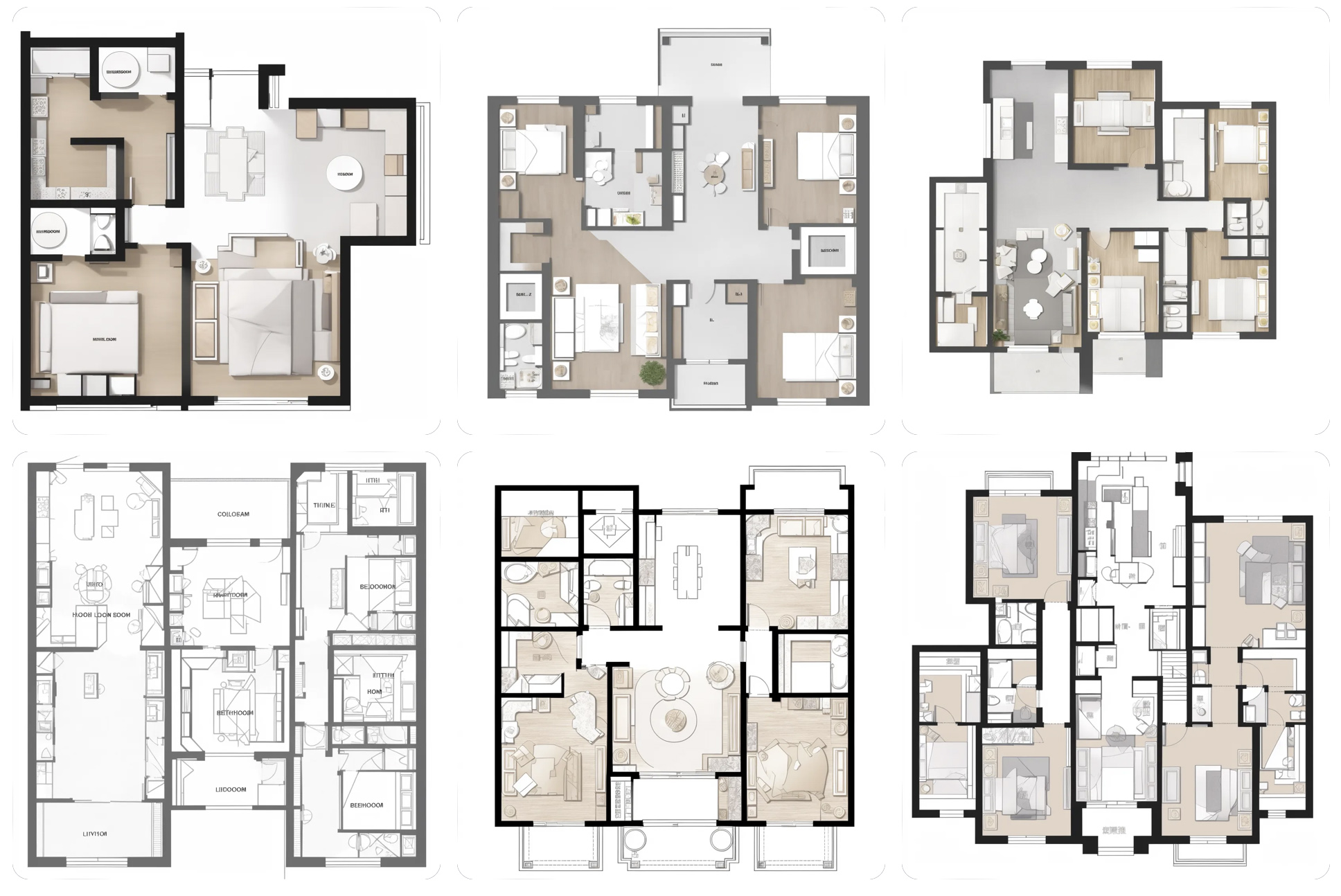}
    \caption{Floorplans-05: Generated by Fine-tuned LoRA weights. }
\end{figure*} 

\begin{figure*}[tbp]
    \centering
    \includegraphics[width=0.99\textwidth]{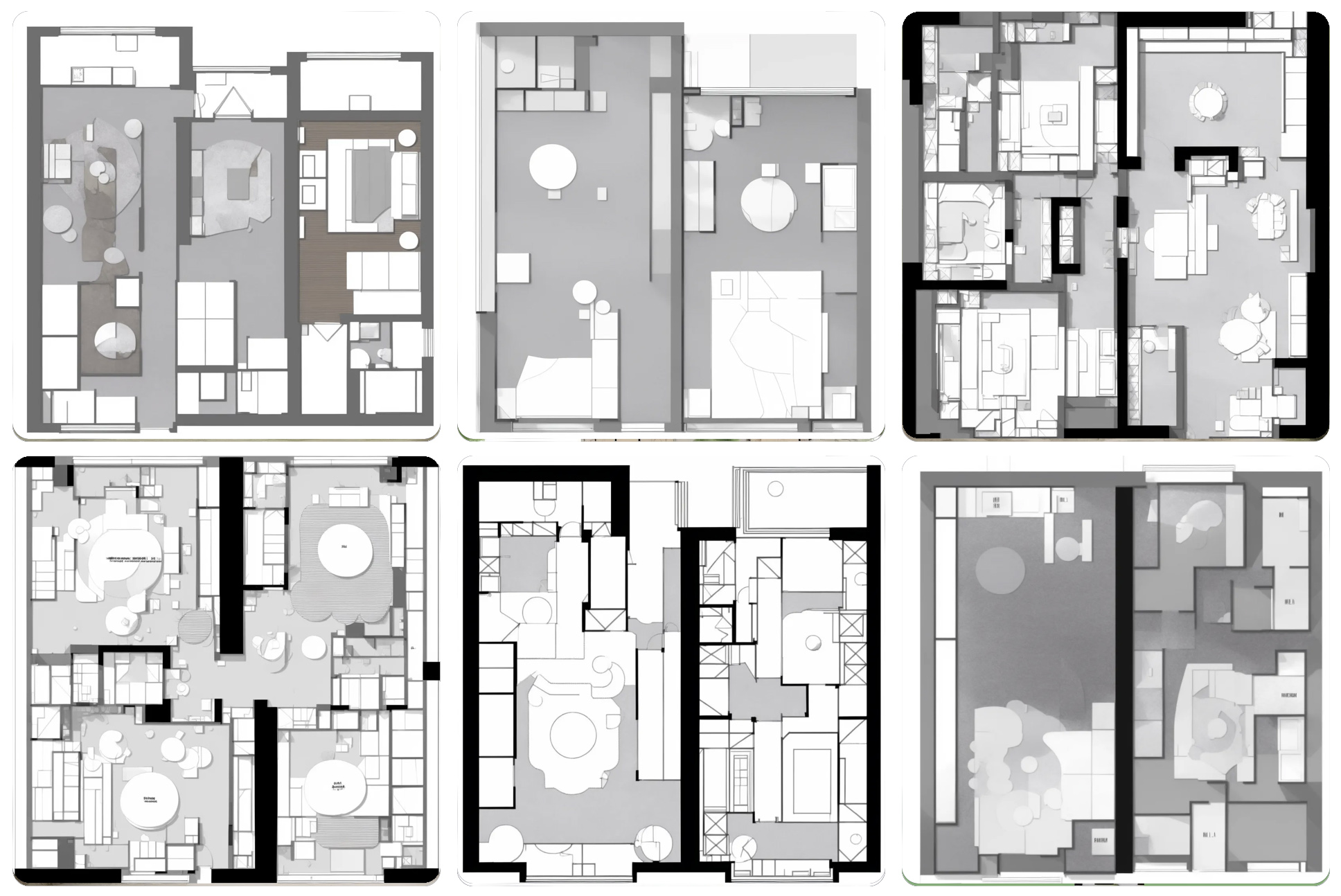}
    \caption{Floorplans-06: Generated by Fine-tuned LoRA weights. }
\end{figure*} 

\subsection{Text Prompts}
\textit{1. house floorplans, modern elegance, a two-story house with an open floor plan; 2. house floorplans, rustic charm, a cabin-style home with a large front porch; 3. house floorplans, urban chic, a compact apartment with efficient use of space; 4. house floorplans, family-friendly, a suburban home with a large backyard; 5. house floorplans, eco-friendly, a sustainable house with solar panels; 6. house floorplans, luxury living, a mansion with multiple bedrooms and a pool; 7. house floorplans, coastal retreat, a beach house with panoramic ocean views; 8. house floorplans, minimalist design, a clean, open space with few furnishings; 9. house floorplans, industrial style, a loft with exposed brick and metal accents; 10. house floorplans, country living, a farmhouse with a wraparound porch; 11. house floorplans, traditional elegance, a classic colonial home with formal dining and living rooms; 12. house floorplans, contemporary style, a sleek, modern house with large glass windows; 13. house floorplans, small space living, a tiny house with a smart layout; 14. house floorplans, urban loft, an open floor plan with high ceilings and large windows; 15. house floorplans, mountain cabin, a cozy home with a stone fireplace; 16. house floorplans, desert oasis, a house with an indoor-outdoor living area; 17. house floorplans, waterfront property, a house with a dock and lake access; 18. house floorplans, historic home, a Victorian house with intricate woodwork; 19. house floorplans, modern farmhouse, a blend of contemporary and rustic elements; 20. house floorplans, beach bungalow, a small, casual house near the shore; 21. house floorplans, luxury apartment, a high-rise unit with city views; 22. house floorplans, suburban comfort, a family home with a large kitchen and playroom; 23. house floorplans, green design, a house with a green roof and energy-efficient features; 24. house floorplans, minimalist loft, a spacious, uncluttered living area; 25. house floorplans, traditional ranch, a single-story home with an open floor plan; 26. house floorplans, lakeside retreat, a house with a large deck overlooking the water; 27. house floorplans, urban townhouse, a multi-level home with a rooftop terrace; 28. house floorplans, mountain lodge, a large home with timber framing and a fireplace; 29. house floorplans, desert modern, a house with clean lines and large windows to capture views; 30. house floorplans, coastal cottage, a charming house with a screened-in porch; 31. house floorplans, contemporary loft, an open, industrial-style space with modern amenities; 32. house floorplans, family estate, a large house with multiple living areas and a guest house; 33. house floorplans, eco-conscious design, a house built with sustainable materials; 34. house floorplans, chic urban flat, a stylish apartment with open living and dining areas; 35. house floorplans, country estate, a large house with a barn and extensive grounds; 36. house floorplans, modern villa, a luxurious house with a pool and outdoor kitchen; 37. house floorplans, rustic lodge, a large house with a stone exterior and wood accents; 38. house floorplans, beach retreat, a house with large windows and a spacious deck; 39. house floorplans, urban studio, a compact living space with an open floor plan; 40. house floorplans, mountain chalet, a cozy house with a sloped roof and fireplace; 41. house floorplans, desert adobe, a house with thick walls and a courtyard; 42. house floorplans, lakeside cabin, a small house with a dock and lake views; 43. house floorplans, historic townhouse, a house with period details and modern updates; 44. house floorplans, modern ranch, a single-story house with an open layout; 45. house floorplans, coastal estate, a large house with ocean views and a guest house; 46. house floorplans, urban loft, a spacious apartment with high ceilings and large windows; 47. house floorplans, family bungalow, a single-story house with a large backyard; 48. house floorplans, green living, a house with solar panels and a green roof; 49. house floorplans, minimalist condo, a sleek, modern living space with an open floor plan; 50. house floorplans, traditional cottage, a charming house with a front porch and garden; 51. house floorplans, luxury penthouse, a high-rise apartment with panoramic city views; 52. house floorplans, suburban retreat, a house with a large yard and pool; 53. house floorplans, eco-friendly home, a house with energy-efficient features; 54. house floorplans, chic loft, a stylish living space with an open floor plan; 55. house floorplans, country mansion, a large house with extensive grounds and a guest house; 56. house floorplans, modern bungalow, a single-story house with a sleek design; 57. house floorplans, rustic farmhouse, a house with a wraparound porch and barn; 58. house floorplans, beach house, a house with a large deck and ocean views; 59. house floorplans, urban apartment, a compact living space with modern amenities; 60. house floorplans, mountain retreat, a house with a stone fireplace and large windows; 61. house floorplans, desert retreat, a house with an indoor-outdoor living area; 62. house floorplans, lakeside home, a house with a dock and panoramic lake views; 63. house floorplans, historic cottage, a charming house with period details; 64. house floorplans, modern estate, a large house with sleek design and luxury features; 65. house floorplans, coastal mansion, a house with ocean views and multiple living areas; 66. house floorplans, urban studio, a compact living space with high ceilings and large windows; 67. house floorplans, family home, a suburban house with a large backyard and playroom; 68. house floorplans, green house, a sustainable house with solar panels and a green roof; 69. house floorplans, minimalist apartment, a sleek living space with an open floor plan; 70. house floorplans, traditional farmhouse, a house with a wraparound porch and garden; 71. house floorplans, luxury condo, a high-rise apartment with city views and modern amenities; 72. house floorplans, suburban estate, a large house with a pool and extensive grounds; 73. house floorplans, eco design, a house built with sustainable materials and energy-efficient features; 74. house floorplans, chic apartment, a stylish living space with an open floor plan; 75. house floorplans, country cottage, a charming house with a front porch and garden; 76. house floorplans, modern house, a sleek, contemporary house with clean lines; 77. house floorplans, rustic retreat, a house with a stone exterior and wood accents; 78. house floorplans, beachside bungalow, a small house with a large deck and ocean views; 79. house floorplans, urban loft, a spacious living area with high ceilings and large windows; 80. house floorplans, mountain house, a cozy house with a sloped roof and fireplace; 81. house floorplans, desert house, a house with thick walls and a courtyard; 82. house floorplans, lakeside retreat, a house with a dock and panoramic lake views; 83. house floorplans, historic home, a house with period details and modern updates; 84. house floorplans, modern ranch, a single-story house with an open layout; 85. house floorplans, coastal villa, a large house with ocean views and a guest house; 86. house floorplans, urban studio, a compact living space with high ceilings and large windows; 87. house floorplans, family retreat, a house with a large yard and pool; 88. house floorplans, eco-friendly design, a house with solar panels and a green roof; 89. house floorplans, minimalist living, a sleek, modern living space with an open floor plan; 90. house floorplans, traditional mansion, a large house with period details and luxury features; 91. house floorplans, suburban bungalow, a single-story house with a large backyard; 92. house floorplans, rustic lodge, a large house with a stone exterior and wood accents; 93. house floorplans, beachfront property, a house with a large deck and ocean views; 94. house floorplans, urban apartment, a compact living space with modern amenities; 95. house floorplans, mountain cabin, a cozy house with a stone fireplace; 96. house floorplans, desert villa, a house with an indoor-outdoor living area; 97. house floorplans, lakeside cabin, a small house with a dock and panoramic lake views; 98. house floorplans, historic estate, a large house with period details and modern updates; 99. house floorplans, modern bungalow, a sleek, single-story house with clean lines; 100. house floorplans, coastal mansion, a house with ocean views and multiple living areas.}

\section{Generating Floorplans from GPT prompts}

The use of GPT-generated prompts to create detailed floorplans through diffusion models stands out. This process leverages the capabilities of two specific models: Stable Diffusion v1.5 and a customized diffusion model trained with Low-Rank Adaptation (LoRA)~\cite{hu2021lora} and specific floorplan styles.

The process begins with generating descriptive text prompts using GPT. These prompts include detailed specifications of the desired floorplan, such as the number of rooms, layout preferences, dimensions, and stylistic elements.

The first diffusion model employed is Stable Diffusion v1.5. This version is well-regarded for its stability and ability to generate high-quality images from textual descriptions. It serves as a robust baseline model, translating the GPT prompts into preliminary floorplan designs.

The second model is a customized diffusion model enhanced with LoRA and trained on a dataset of various floorplan styles. LoRA allows the model to adapt to specific patterns and styles with fewer parameters, making it highly efficient. This model further refines the initial designs provided by Stable Diffusion v1.5, incorporating more nuanced details and stylistic elements that align with the specific requirements.

As shown in Figure 5, 6 and 7, the LoRA model fine-tuned with specific styles is able to generate floorplans that better match design requirements, while the initial Stable Diffusion model produces comparatively subpar results (As shown in Figure 2, 3 and 4). By comparing the two approaches, we can clearly observe the superiority of the LoRA model in terms of detail and stylistic accuracy. While Stable Diffusion can quickly generate a rough layout, it falls short in the details and design consistency. In contrast, the LoRA model, fine-tuned with specific styles, better captures the design requirements, producing floorplans that are not only more precise in layout but also more consistent with the expected stylistic details.

Through the comparative study of these two models, we found that the LoRA model has significant advantages in generating complex floorplans. Although the Stable Diffusion model is faster, it often struggles with complex design elements and details. This is mainly because the Stable Diffusion model was not specifically optimized for particular styles during training, making it difficult to fully meet the specific design style requirements during generation.

The LoRA model, on the other hand, stands out due to its fine-tuning with specific styles. This enables the LoRA model to better adapt to design requirements, capturing subtle stylistic elements. The generated floorplans are not only more aesthetically pleasing but also more functional. One key advantage of this approach is its flexibility, allowing adjustments according to different design needs, thus producing more personalized and customized floorplans. Fine-tuning the LoRA model with specific styles significantly improves the quality of floorplan generation, meeting higher design requirements and detail precision. This method has broad application prospects in fields such as architectural design and interior decoration.

\section{Conclusion}

AI-driven text-to-image synthesis is transforming architectural design by converting textual descriptions into detailed and functional floor plans. We has examined the technical aspects, practical methods, and future potential of AI-assisted floor plan design, focusing on tools like Stable Diffusion. Our study evaluated these AI models' effectiveness in generating residential floor plans from various inputs, highlighting their strengths and limitations. The findings suggest that architects can leverage AI tools to streamline design processes, generate diverse design options, and foster creativity and collaboration. By demonstrating AI's capability to produce high-quality, functional floor plans, this research underscores AI's potential to enhance efficiency and innovation in architectural design.

\bibliographystyle{ACM-Reference-Format}
\bibliography{sample-base}

\appendix

\end{document}